# Evaluation of the efficacy of RUTI and ID93/GLA-SE vaccines in tuberculosis treatment: in silico trial through UISS-TB simulator

Giulia Russo*
Dept. of Drug Sciences
University of Catania
Catania, Italy
giulia.russo@unict.it

Francesco Pappalardo*
Dept. Of Drug Sciences
University of Catania,
Catania, Italy
francesco.pappalardo@unict.it

Miguel A. Juarez
School of Mathematics & Statistics
University of Sheffield
Sheffield, UK
m.juarez@sheffield.ac.uk

Marzio Pennisi
Dept. of Mathematics and Computer Science
University of Catania
Catania, Italy
mpennisi@dmi.unict.it

Pere Joan Cardona
Archivel Farma S.L.
Badalona, Spain
pj.cardona@gmail.com

Rhea Coler
Infectious Disease Research Institute
Seattle, USA
rhea.coler@idri.org

Epifanio Fichera
Etna Biotech s.r.l.
Catania, Italia
epifanio.fichera@etnabiotech.it

Marco Viceconti
Dept. of Department of Industrial Engineering
University of Bologna
Bologna, Italy
marco.viceconti@unibo.it

Corresponding author: Marco Viceconti

***Abstract*— Tuberculosis (TB) is one of the deadliest diseases worldwide, with 1,5 million fatalities every year along with potential devastating effects on society, families and individuals. To address this alarming burden, vaccines can play a fundamental role, even though to date no fully effective TB vaccine really exists. Current treatments involve several combinations of antibiotics administered to TB patients for up to two years, leading often to financial issues and reduced therapy adherence. Along with this, the development and spread of drug-resistant TB strains is another big complicating matter. Faced with these challenges, there is an urgent need to explore new vaccination strategies in order to boost immunity against tuberculosis and shorten the duration of treatment. Computational modeling represents an extraordinary way to simulate and predict the outcome of vaccination strategies, speeding up the arduous process of vaccine pipeline development and relative time to market. Here, we present EU - funded STriTuVaD project computational platform able to predict the artificial immunity induced by RUTI and ID93/GLA-SE, two specific tuberculosis vaccines. Such an in silico trial will be validated through a phase 2b clinical trial. Moreover, STriTuVaD computational framework is able to inform of the reasons for failure should the vaccinations strategies against M. tuberculosis under testing found not efficient, which will suggest possible improvements.***

***Keywords*—Tuberculosis, vaccine, in silico clinical trials, simulation**

## I. Introduction

Tuberculosis (TB) is one of the top 10 causes of death around the globe and killed 1.7 million people in 2016, according to the World Health Organization [1]. Spread through the air, it takes just a sneeze or cough to diffuse from one person to another one [2]. Most fatalities occur in poorer countries even though no population today is immune or isolated from the risk to be affected by TB [3]. To date non fully effective TB vaccines exists and, despite being both preventable and curable, it can be difficult for TB infected patients to get live-saving care [4]. Current treatment can involve antibiotics administration for up to two years, potentially becoming a financial and social burden and resulting in patients stopping their medication [5]. At times,

* The authors wish it to be known that the first two authors should be regarded as joint first author.

TB is not diagnosed and dealt with the development and spread of drug-resistant strains [6], [7].

Faced with these challenges, several EU-funded projects are working to identify and maximize the potentiality of novel vaccines in order to save lives, stem the spread of infection and shorten treatment. In this context, the HORIZON 2020-STriTuVaD project is leading the way to fight against TB through the development of an innovative technology named Universal Immune System Simulator (UISS), based on a computational modeling infrastructure able to simulate and predict the outcome of two specific vaccination strategies, i.e., RUTI vaccine and ID93+GLA-SE vaccine [8], [9] along with the relevant individual human physiology and physiopathology of patients affected by Mycobacterium tuberculosis (MTB).

To this aim we generated a set of digital populations of individuals providing a reliable prediction of phase III outcomes on the basis of the data collected during the phase II clinical trial, as agreed in the project.

To analyze the behavior of possible therapeutic interventions, we implemented mechanisms of action (MoA) of ID93+GLA-SE vaccine in our pre-existing version of UISS-TB computational platform, in which RUTI vaccine was previously implemented [10], along with the entire dynamics of MTB and interactions with the immune system machinery. We analysed the associated immune response induced by both vaccines and we applied UISS-TB computational model to set up a library of digital TB patients, through the identification of a “vector of features” that combines biological and pathophysiological parameters of tuberculosis patients.

In this scenario, our simulation infrastructure was also able to reveal few not-responders patients after TB vaccines administration.

## II. Materials and Methods

### A. *Universal Immune System Simulator (UISS)*

The computational platform is represented by the Universal Immune System Simulator (UISS), a multi-scale (cellular and molecular level), multi-compartment, polyclonal agent based simulator of the immune system dynamics [11]. Cellular entities can take up a state from a certain set of suitable states and their dynamics is realized by means of state changes. They include lymphocytes (i.e. B lymphocytes, helper, cytotoxic and regulatory T lymphocytes and natural killer cells) and monocytes (i.e. macrophages (M) and dendritic cells). For what concerns molecules, the model distinguishes between simple small molecules like interleukins or signaling molecules in general and more complex molecules like immunoglobulins and antigens. At the same level of entities, UISS implements immune system activities. They include both interactions and functions, referring respectively to the main immune system tasks. Moreover, UISS represents receptors and ligands as bit strings and use a string matching rule to model affinity. In particular, specificity is implemented in UISS by a bit-string polyclonal lattice method [12].

In this context, UISS was extended to consider immune system dynamics at a large scale. This was achieved extending UISS platform in three critical points: *i)* the peripheral blood compartment increased in size in order to deal with a representative portion of peripheral blood and included immune system entities circulating through it; *ii)* immune system repertoire implementation in order to take into account an immune system potential diversity at human natural scale i.e. about $10^{20}$ order magnitude of T and B cells clonotypes; *iii)* compartments addition to simulate critical organs targeted by tuberculosis i.e., lungs and near lymph nodes.

UISS web-interface is developed using the Flask micro-server environment and Python programming language. Figure 1 shows a screenshot of the web Graphic User Interface (GUI) in which it is possible to set different combinations of cellular and non-cellular parameters. The simulation will be performed asynchronously. This allows to launch the process separately from the main thread, notifying the user when the simulation is finished and performing multiple simulations simultaneously.

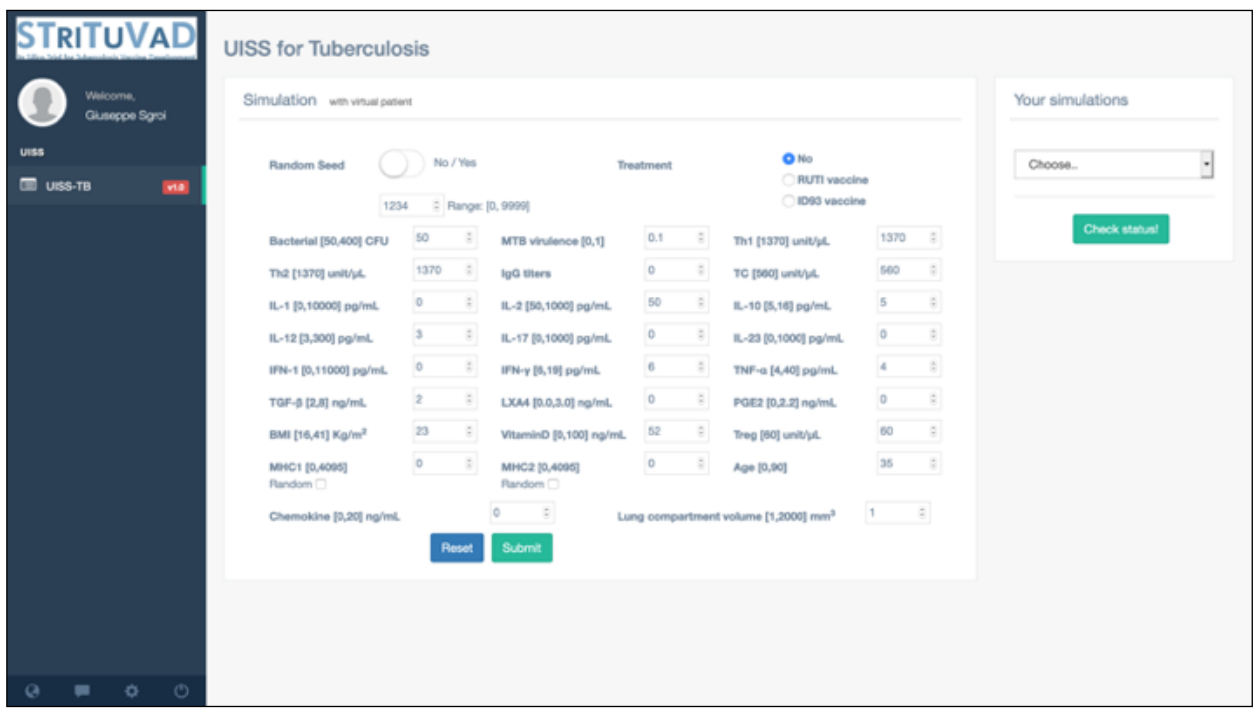

Figure 1. UISS web-GUI. On the left, one can see a set of biological and physiopathological parameters, referring to the vector of features created for the personalization of TB digital patients. On the right, one can see a box called "your simulations", containing the list of all the simulations, sorted by their creation date and classified in “in execution” or in “completed” status.

### B. *TB vaccines*

To analyze the behavior of a possible therapeutic intervention, we simulated through our modeling infrastructure the two vaccines we are going to test inside the STriTuVaD project i.e., RUTI (developed by Archivel Farma S.L.) and ID93/GLA-SE (developed by Infectious Disease Research Institute (IDRI)) vaccines. To this end, we implemented inside UISS ID93/GLA-SE mechanism of action (MoA).

For the sake of completeness, RUTI is a polyantigenic liposomal vaccine made of detoxified, fragmented M. tuberculosis cells (FCMtb), while ID93/GLA-SE consists of a fusion of four M. tuberculosis proteins: Rv1813, Rv2608, Rv3619, and Rv3620. The vaccination strategy includes also the GLA-SE adjuvant.

We modeled the immune response induced by one boost of 25ug dosage of RUTI and two boosts of 100ng, with the second dosage two months later after the first one for ID93/GLA-SE, accordingly to the clinical trial protocol as agreed in the STriTuVaD project.

## C. *Generation of libraries of digital patients*

To reproduce biological diversity of TB patients, an appropriate procedure for the generation of libraries of digital patients has been developed. This has been achieved through the implementation of three specific strategies: *i)* the creation of the initial immune system repertoire in a stochastic way taking also into account the DNA fragments assembly and Class I or Class II HLA patterns; *ii)* the simulation of immune system interactions; *iii)* the identification of a "vector of features" that associates both biological and pathophysiological parameters that personalize the digital patient and reproduce the physiology and the pathophysiology of TB patients.

The first is the creation of the initial immune system repertoire, generated in a way that simulates the DNA fragments assembly accounting for the inherent stochasticity of the process, but also on the presence of the Class I or Class II HLA patterns. The second is related to the simulation of immune system interactions. These take place in a different order, according to a sequence that depends on a series of events whose generated from a selection of a random seed. This imitates, to some extent, the randomness of the immune response in the initial phases, where innate immunity takes place (consider, for example, the case that a fully matching CD4 T cell is present at the specific lymph node where a DC is presenting the processed antigen). The last approach is represented by the identification of a "vector of features" that defines a specific patient through a vector of 26 features as follows: 1) drug Sensitive (DS)/multi-drug resistant (MDR); 2) bacteria Load (BL) in sputum; 3) MTB strain; 4) CD4-Th1; 5) CD4-Th2; 6) IgG titers; 7) CD8 T cells; 8) IL-1; 9) IL-2; 10) IL-10; 11) IL-12; 12) IL-17; 13) IL-23; 14) IFN Type I: 15) IFN-γ; 16) TNF-α; 17) TGF-β; 18) LXA4; 19) PGE2; 20) Chemokines; 21) Vitamin D; 22) HLA-1; 23) HLA-2; 24) FoxP3; 25) Age; 26) BMI.

# III. RESULTS

We run a total of 60 simulations i.e., 30 for digital in silico patients treated with RUTI vaccine and 30 for digital in silico patients treated with ID93/GLA-SE. Simulation results of UISS-TB applied to the sample set of digital in silico patients are shown in the following figures. We present, for each biological entity, both the mean behavior and the +/- SD (blue lines).

The average effects of the vaccination based on RUTI vaccine and the ones based on ID93/GLA-SE vaccine are shown, taking into account the alveolar macrophages (AM) dynamics, CD4 Th1, IFN-γ and immunoglobulins levels. Figure 2 shows the alveolar macrophage dynamics after RUTI administration accordingly to one of the approved protocol. We can observe that the average necrotic AM population is considerably reduced indicating an effective immune response elicited by RUTI vaccine, decreasing the probability of disease reactivation.

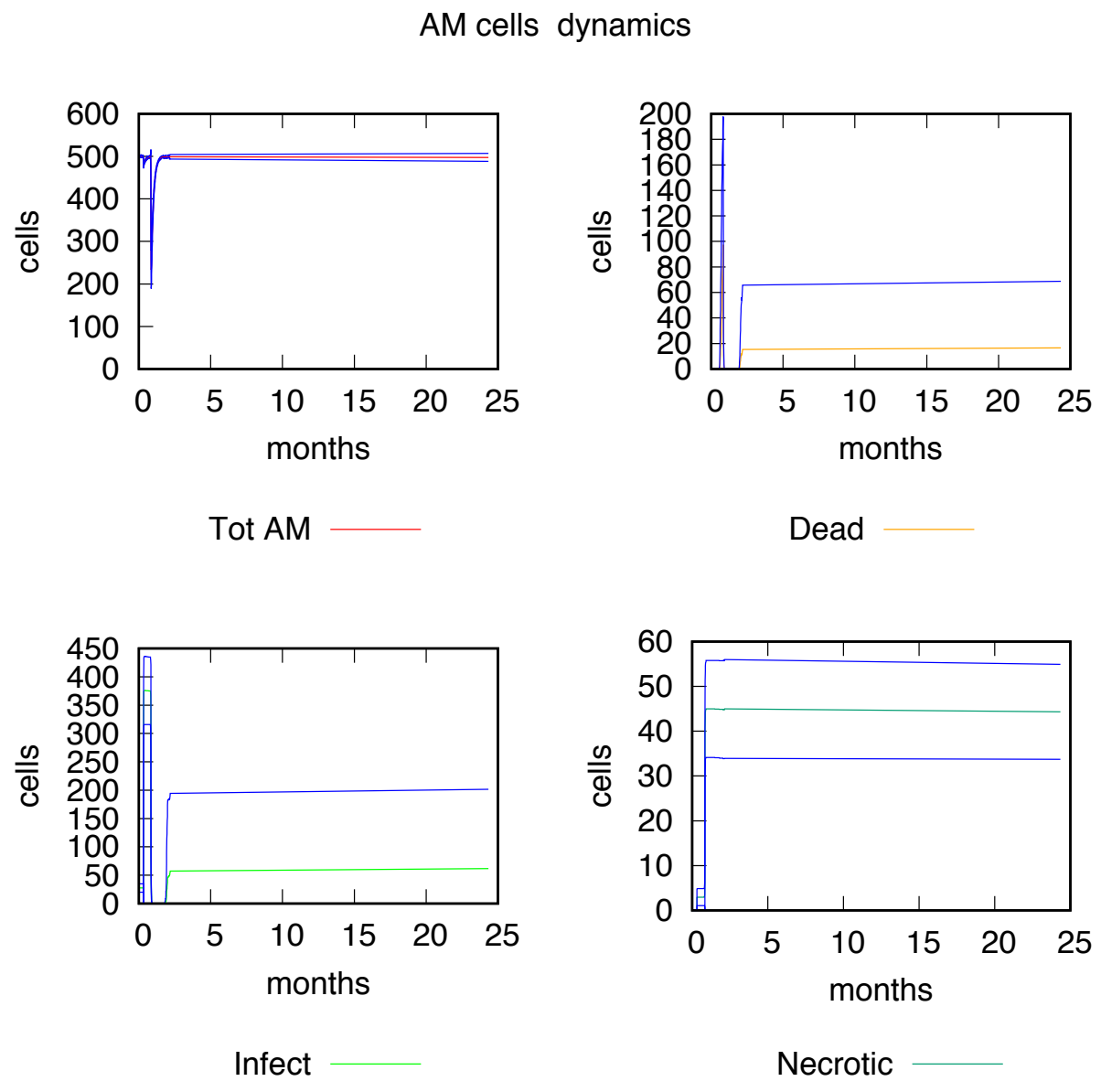

Figure 2. AM population detailed dynamics with RUTI vaccine administration.

Figure 3 shows the alveolar macrophage dynamics after ID93/GLA-SE administration. We can observe that the average necrotic AM population is considerably reduced indicating an effective immune response elicited by this vaccination strategy, decreasing the probability of disease reactivation.

Then, in figure 4, a strong Th1 response is induced with a down-regulation of Th2 response, with the induction of immunological memory, after RUTI administration, while in figure 5 high levels of IFN-γ are present, in good agreement with the results presented in specific literature.

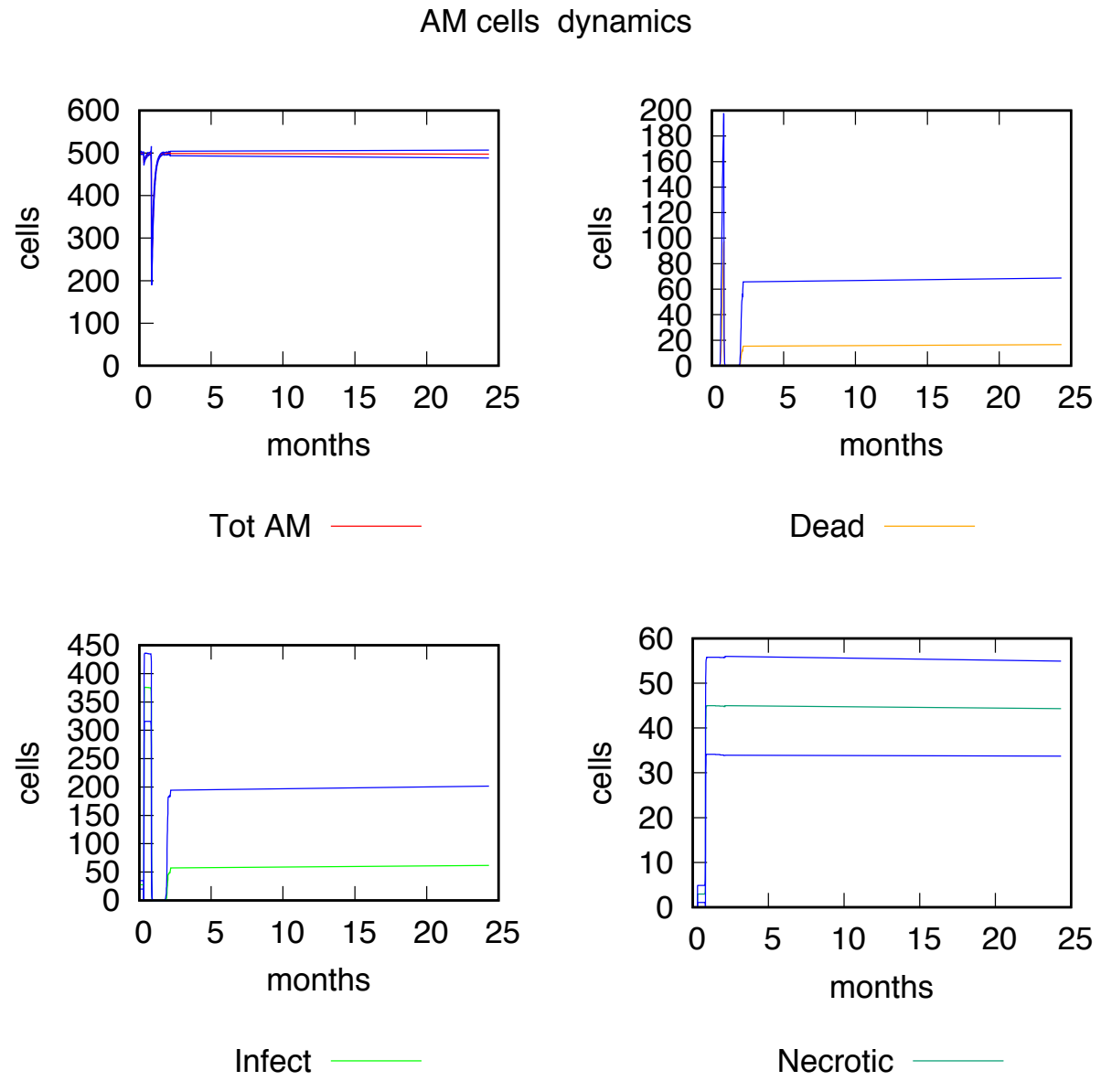

Figure 3. AM population detailed dynamics with ID93/GLA-SE vaccine administration.

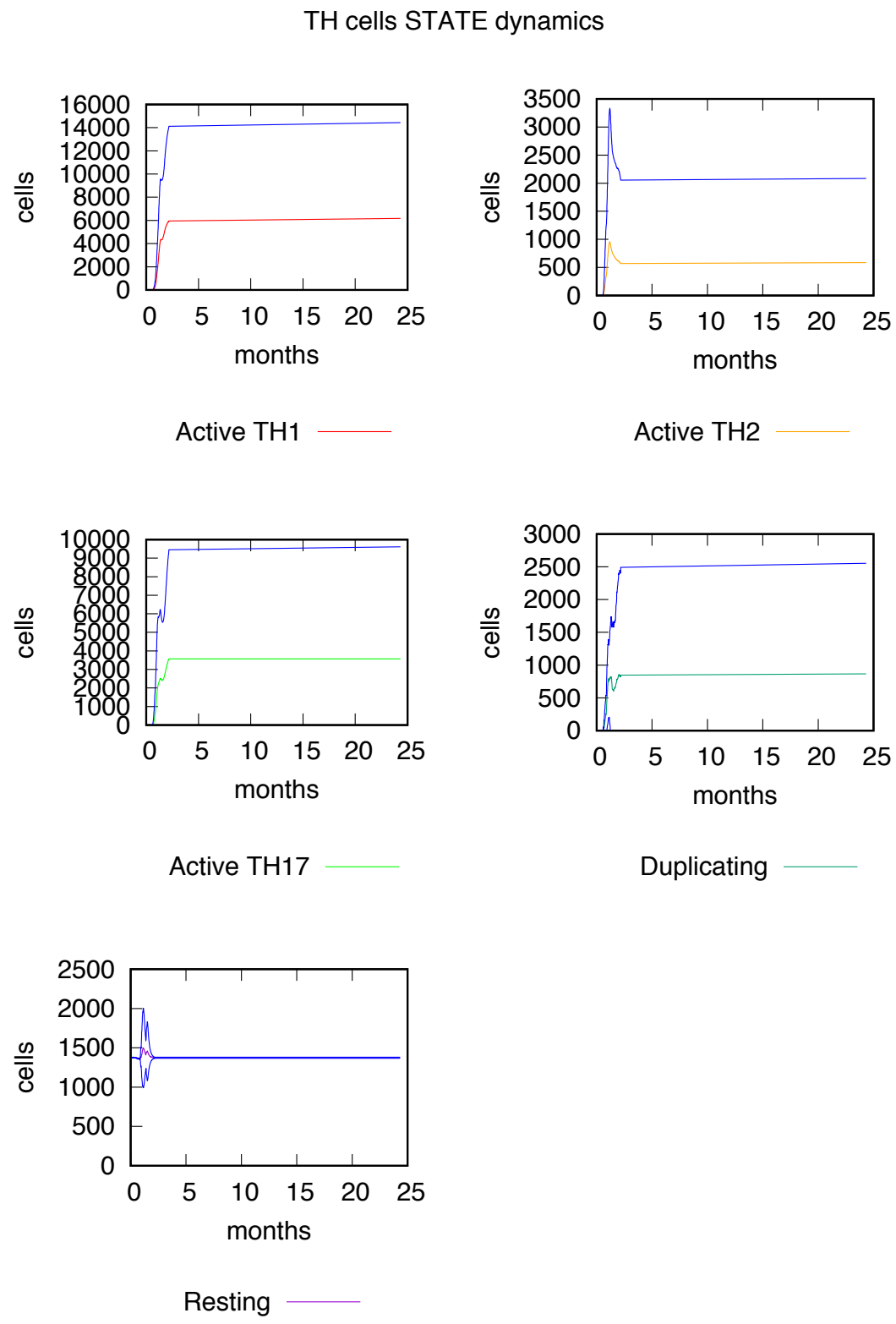

Figure 4. CD4 T cell population detailed dynamics after RUTI vaccine administration.

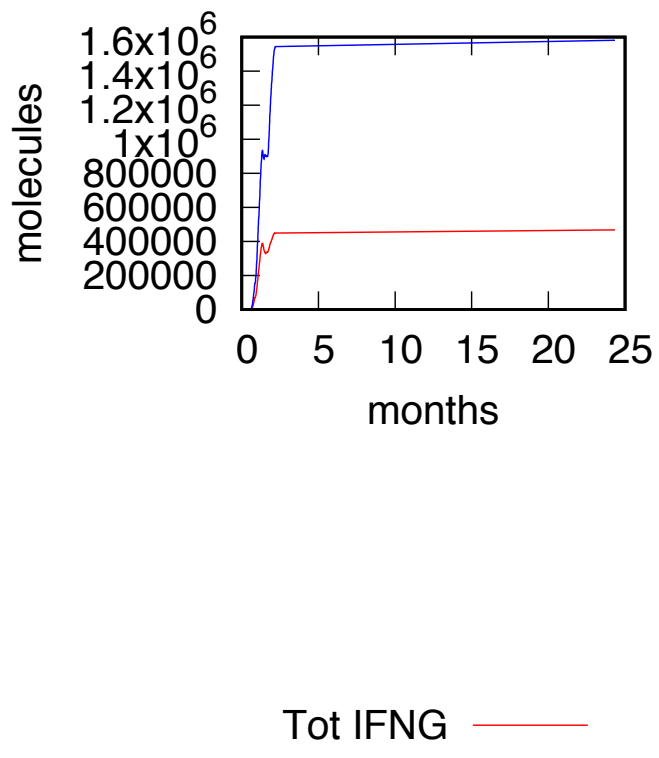

Figure 5. IFN-γ levels after RUTI administration.

Then, after ID93/GLA-SE administration, a strong Th1 response is induced with a down-regulation of Th2 response, along with the induction of immunological memory as depicted in figure 6, while in figure 7 high levels of IFN-γ are present, in good agreement with the results presented in specific literature.

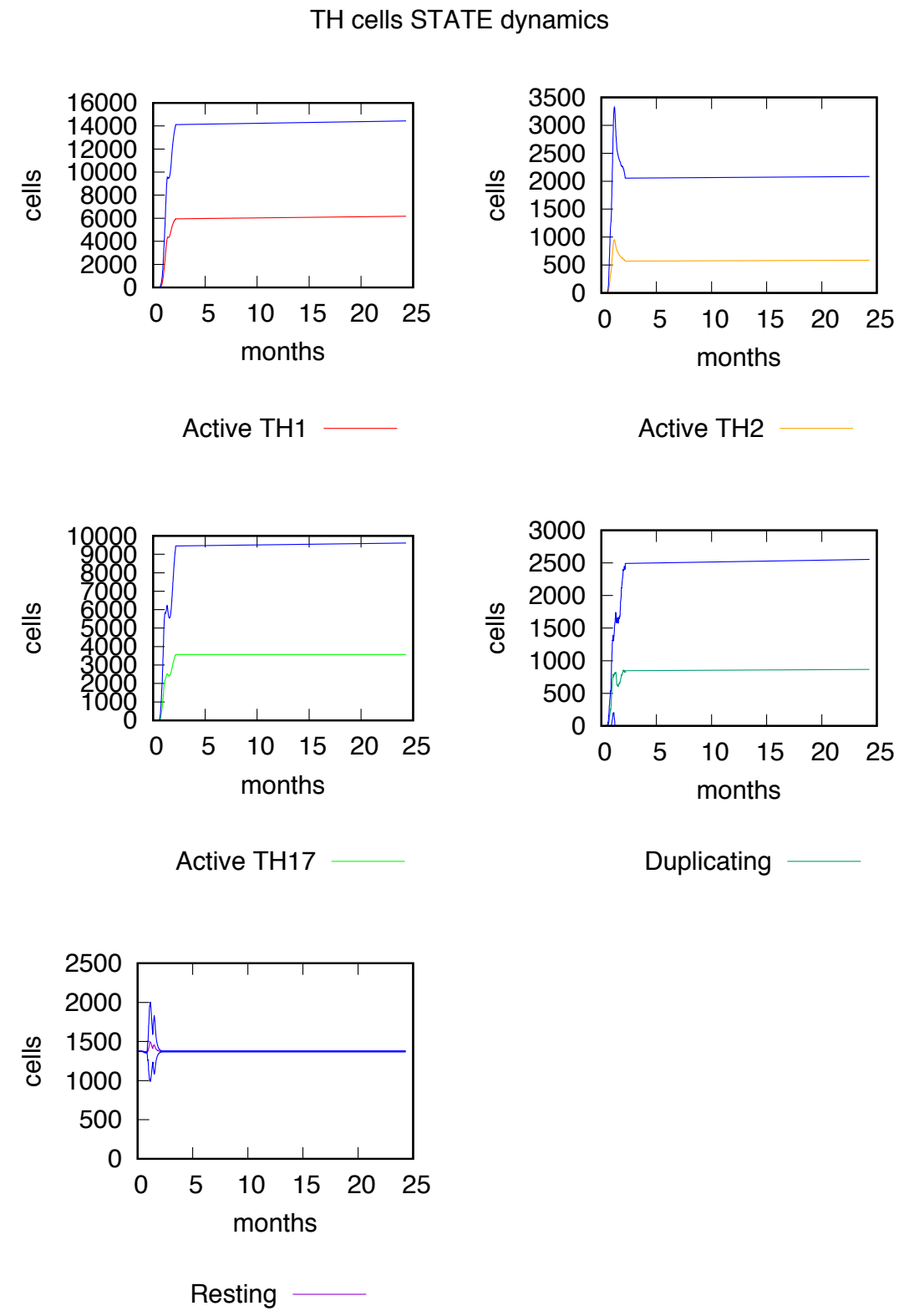

Figure 6. CD4 T cell population detailed dynamics after ID93/GLA-SE vaccine administration.

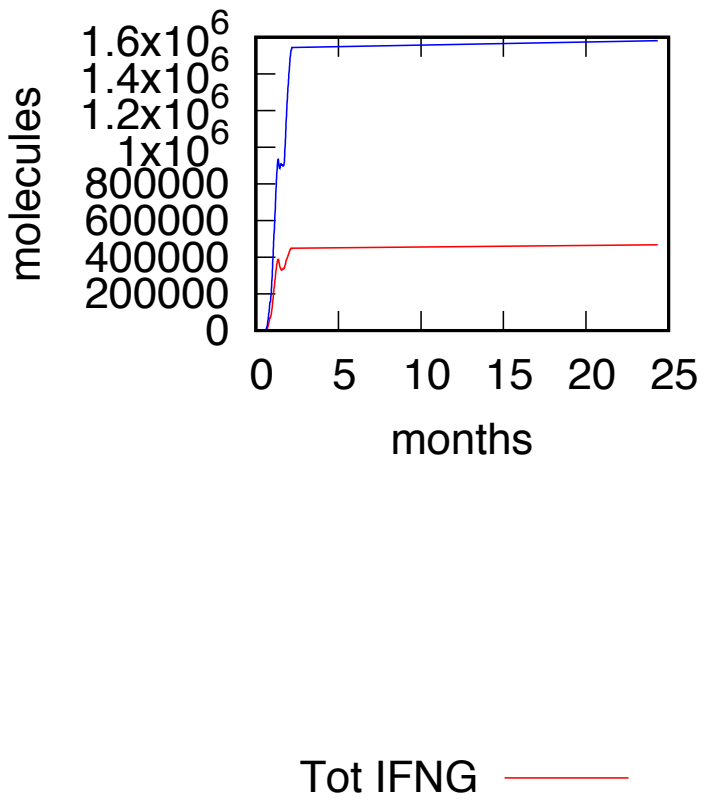

Figure 7. IFN-γ levels after ID93/GLA-SE administration.

It is worth to mention that, looking at the +/- SD data, UISS-TB is able to identify few digital in silico patients that are actually not responding to vaccines stimuli. As an example, we reported in figures 8 and 9 the AM population dynamics, CD4 T cell population dynamics and IFN-γ levels of a specific case of not responder, respectively for RUTI and ID93/GLA-SE vaccine.

Summarizing, UISS-TB reveals that not-responders patients are identified by insufficient CD4 T cell Type 1 response along with the correspondent low levels of IFN-γ. This could depend on specific patient immune system repertoire that do not allow an efficient antigen presentation process followed by an impaired CD4 T cell response.

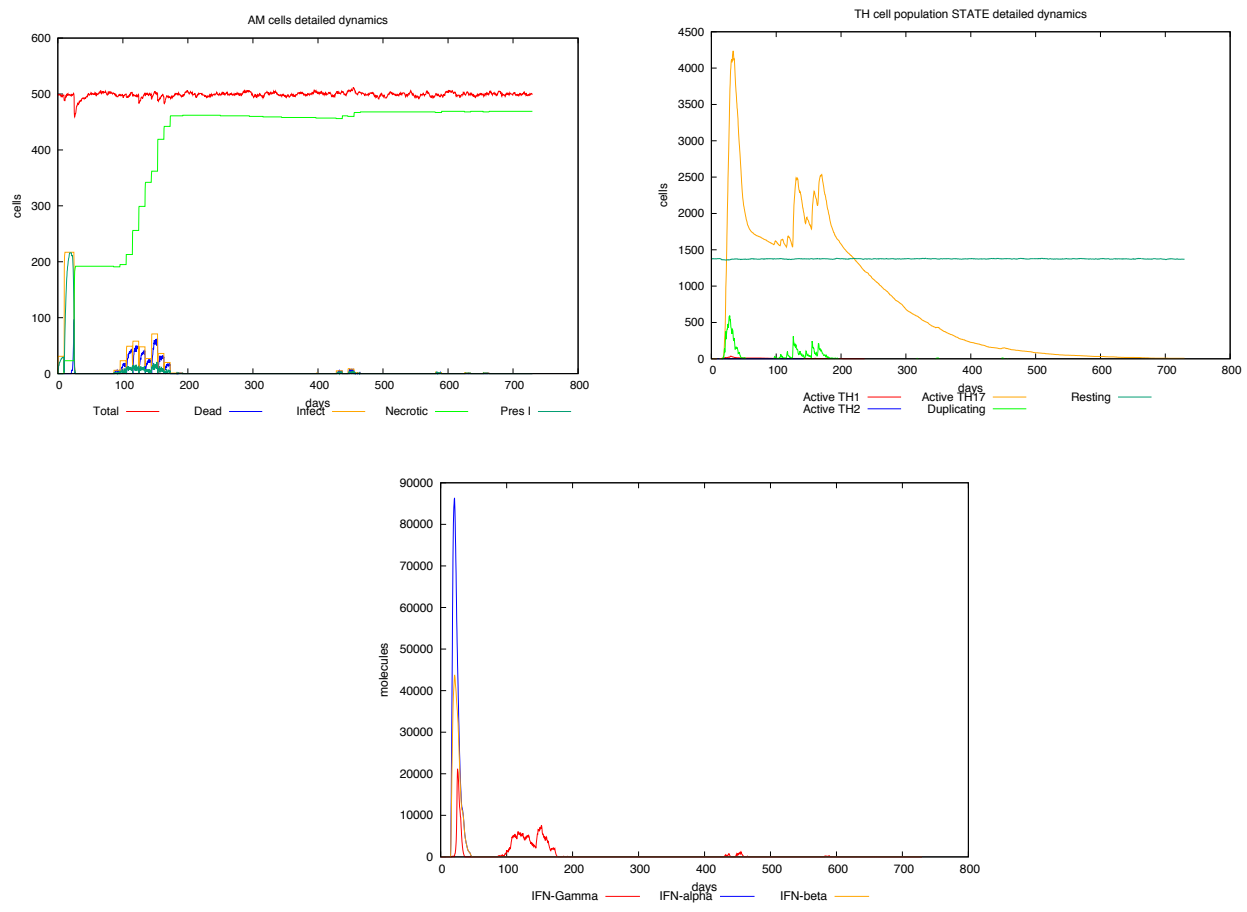

Figure 8. AM population dynamics, CD4 T cell population dynamics and IFN-gamma levels in a not responder case after RUTI administration.

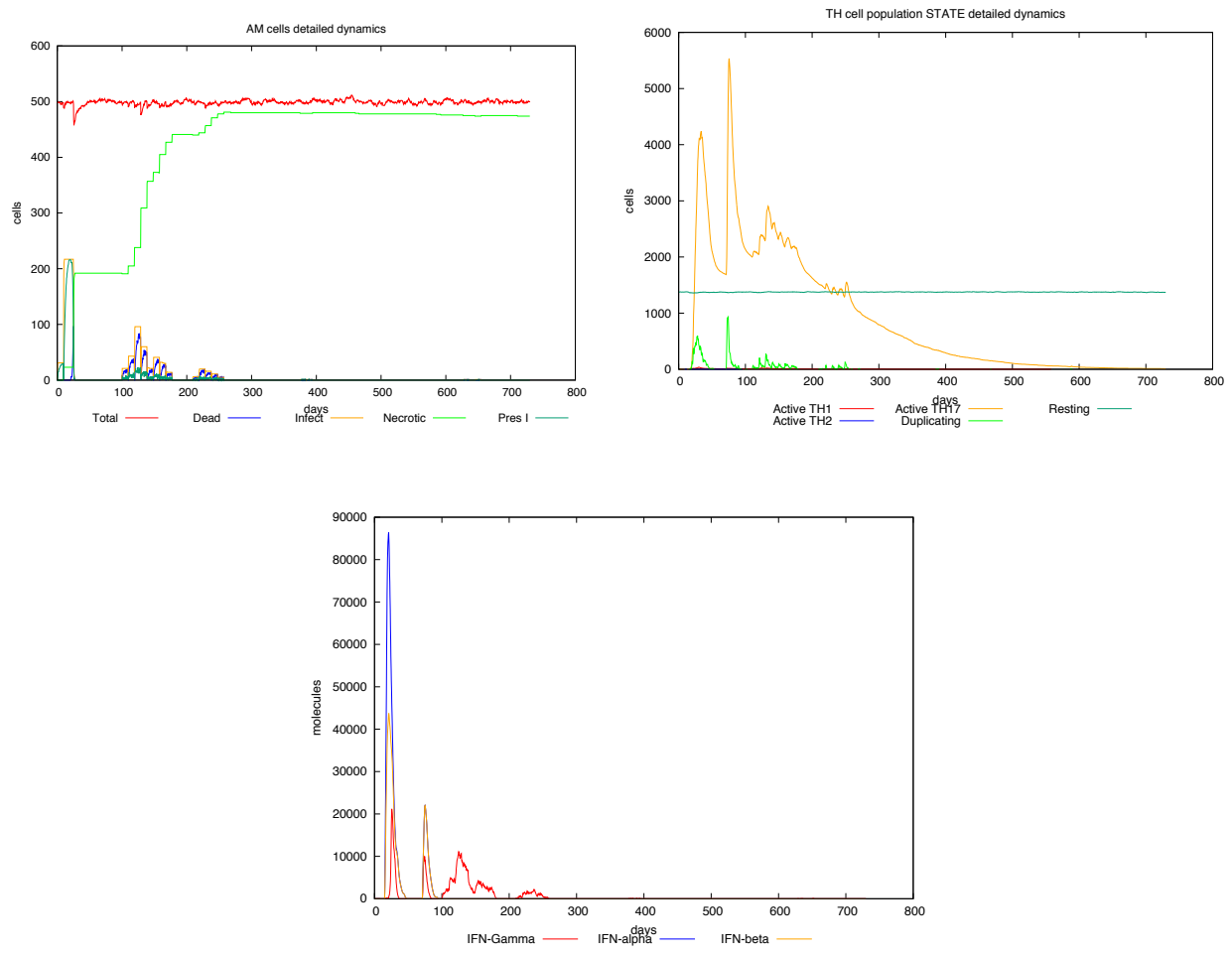

Figure 9. AM population dynamics, CD4 T cell population dynamics and IFN-gamma levels in a not responder case after ID93/GLA-SE administration.

## IV. Conclusions

UISS-TB, the simulation platform employed in STriTuVaD project, has revealed a very good ability to capture the essential immune system responsiveness elicited by two specific vaccination strategies against tuberculosis disease. Moreover, UISS was able to identify few cases of bad responders accordingly to the specificity of CD4 T cell dynamics. This could represent a strong evidence for a successful clinical trial.

## Acknowledgment

Authors of this paper acknowledge support from the STriTuVaD project. The STriTuVaD project has been funded by the European Commission, under the contract H2020-SC1-2017-CNECT-2, No. 777123. The information and views set out in this article are those of the authors and do not necessarily reflect the official opinion of the European Commission. Neither the European Commission institutions and bodies nor any person acting on their behalf may be held responsible for the use which may be made of the information contained therein.